\documentclass[a4paper,11pt]{article}

\usepackage{jcappub}
\usepackage{url}
\usepackage{bm}
\usepackage{amsmath}
\usepackage{appendix}

\def\aap{A\& A}

\def\apj{ApJ}

\def\apjs{ApJS}

\def\mnras{MNRAS}

\title{The excursion set model a step beyond :Environmental dependence}

\author[a]{Nicos Hiotelis}

\affiliation[a]{1st  Lyceum of Athens, Ipitou 15, Plaka, 10557,
Athens,  Greece}

\emailAdd{hiotelis@ipta.demokritos.gr}

\abstract{In terms of the excursion set model, we used Monte Carlo methods in order to study the non-Markovian stochastic evolution of the smoothed overdensity $\delta$ at scale $S$. For  a Gaussian density field, smoothed by the  top-hat filter, in real space, we used random walks, with the correct correlation between scales, in order to calculate constrained probabilities  and to connect the number of structures with the overdensity of their environment. A constant barrier is used, lower than the usual one but which improves the consistency of our results with those of N-body simulations.We present comparisons of multiplicity functions with those resulting from N-body simulations as well as number densities of descendant haloes and of their progenitors and we connect them to the density of their large environment.There exists a dependence of the number of haloes on their environment which is different for different masses of these haloes.This number increases for larger overdensity of the environment for large haloes and decreases for small ones. The number of  progenitors of these haloes is insensitive to the overdensity of their large environment.
}


\begin{document}

\maketitle

\flushbottom

\section{Introduction}
Several studies in the literature investigate the dependence of galaxy properties on their environment. Some older ones detect no dependence of halo clustering on properties such as concentration or formation time ( \cite{Percival2003},\cite{Zentner2005}), while others show that haloes in dense regions are formed at slightly earlier times than haloes of the same mass in less dense regions (\cite{Sheth2004}). On the other hand, more recent studies yield that a variety of characteristics of haloes depend on their environment  (\cite{Gao2007}, \cite{Zentner2014}).
However, it is interesting to try to connect the properties of haloes with their environment, a problem that can be studied by various methods. Among these methods N-body simulations are the most powerful, while analytical or semi-analytical ones have their own advantages, being more economic and shedding more light on the process of structure formation.\\
In this paper, we use the excursion set formalism  based on the ideas of \cite{Press1974}(PS).The initial approach of PS is a very simple model which is based on the assumptions of Gaussian density perturbation and the spherical collapse model. This approach overpredicts the number of objects of small masses and underpredicts the number of those at small masses (e.g. \cite{Jenkins2001}; \cite{White2002}). These problems are not solved even for the extended PS-formalism of  \cite{Peacock1990}, \cite{Bower1991},\cite{Bond1991}, \cite{Lacey1993},\cite{Gardner2001}.\\
 In this study, we  use a non-Markovian stochastic process in order to find first crossing distributions of a constant barrier and we study the assembly history of the haloes which are formed according to the excursion set idea. We assume a Gaussian density field with a correlation between scales that matches the correlation which results from a $\Lambda$CDM power spectrum and a top-hat filter in real space. We study the resulting unconstrained and constrained mass functions and we compare the results with those of N-body simulations.\\
The paper is organized as follows: In Section 2, we discuss the excursion set approach as well as the stochastic process used, and in Section 3 we give details about the numerical methods used as well as the results about unconstrained and constrained mass functions. Comparison with the results available from N-body simulations is also made. Finally a short discussion is given in Sect.4.

\section{The stochastic process.}
In this section, we give a short review of the excursion set model and  describe the stochastic process presented in \cite{Hiotelis2017} which is used in this study.\\
 The smoothed density perturbation  of a spherical region of radius $R$ is
 \begin{equation}
 \label{a00}
 \delta(R)=\int W_f(r;R)\hat{\delta} (r)4\pi r^2\mathrm{d}r
 \end{equation}
   where $\hat{\delta} (r)$ is the density at distance $r$ from the center of the spherical region and $W_f$ is a smoothing filter. Reducing  $R$, the variable $\delta$ executes a random walk that depends on the form  of the density field and on  the smoothing filter $W_f$. If the density field is Gaussian, then $\delta$ is a central Gaussian variable and the probability of being between $x$ and $x+\mathrm{d}x$  is given by
 \begin{equation}
  \label{a0}
  p(\delta(R)=x)\mathrm{d}x=\frac{1}{\sqrt{2\pi \sigma^2(R)}}\mathrm{exp}\left[-\frac{x^2}{2\sigma^2(R)}\right]\mathrm{d}x
  \end{equation}
   For a spherically symmetric filter, the variance at radius $R$ is given by
  \begin{equation}
  \label{a1}
  S(R)\equiv \sigma^2(R)=\frac{1}{2\pi^2}\int_{0}^{\infty}k^2P(k){\widehat{W}_f}^2(k;R)\mathrm{d}k
  \end{equation}
  where $\widehat{W}_f$ is the Fourier transform of the filter and $P$ is the power spectrum.\\
  The correlation of values of $\delta$ between scales is given by the autocorrelation function that is
  \begin{equation}
  \label{a2}
  \langle\delta(R)\delta(R')\rangle=\frac{1}{2\pi^2}\int_{0}^{\infty}k^2P(k){\widehat{W}_f}(k;R){\widehat{W}_f}(k;R')\mathrm{d}k
  \end{equation}
  According to hierarchical scenario of structure formation in the Unverse, $S$ is a decreasing  function of $R$ and obviously $R$  is an increasing function the the mass $M$ contained in the sphere of radius $R$, then, $S$ is a decreasing function of mass. \\
  It is also obvious the significant role of the filter since its shape defines the correlation between scales.
  The most interesting filter, in terms  of its obvious physical meaning, is the top-hat in real space given by
  \begin{equation}
  \label{a3}
  W_f(r;R)=H\left(1-\frac{r}{R}\right)\frac{1}{\frac{4}{3}\pi R^3}
  \end{equation}
  where $H$ is the Heaviside step function. The Fourier transform of the filter is given by,
  \begin{equation}
  \label{a4}
  \widehat{W}(k;R)=\frac{3[\mathrm{sin}(kR)-kR\mathrm{cos}(kR)]}{k^3R^3}
  \end{equation}
 In this paper, we assume a Gaussian density field and the top-hat filter in real space. We use a flat model for the Universe with
  present day density parameters $\Omega_{m,0}=0.3$ and
   $ \Omega_{\Lambda,0}\equiv \Lambda/3H_0^2=0.7$, where
  $\Lambda$ is the cosmological constant and $H_0$ is the present day value of Hubble's
  constant. We have used the value $H_0=100~\mathrm{hKMs^{-1}Mpc^{-1}}$
  and a system of units with $m_{unit}=10^{12}M_{\odot}h^{-1}$,
  $r_{unit}=1h^{-1}\mathrm{Mpc}$ and a gravitational constant $ G=1$. At this system of units
  $H_0/H_{unit}=1.5276.$ Regarding  the power spectrum, we  employed the $\Lambda CDM$ formula proposed by
  \citep{Smith1998}.\\
 It is shown in  \citep{Maggiore2010a} that for the above filter, the correlation between scales is well approximated  by the relation
 \begin{equation}
 \label{cor}
  <\delta(S)\delta(S')>=S'+\lambda\frac{S'(S-S')}{S}
 \end{equation}
 for $S'<S$ with $\lambda$ about 0.45. This conclusion is also derived by \cite{Hiotelis2017}(see fig.1 therein).\\
 The purpose of the excursion set model is to map the initial density field of the Universe to the number of structures (haloes) of various masses at
 any epoch. Let consider a point $A$ in the early Universe. We use \eqref{a00} to calculate the mean overdensity for a sphere of radius $R$ and center $A$. Since the Universe is homogeneous at large scales $\delta\rightarrow 0$ as $R\rightarrow +\infty$ or equivalently $S\rightarrow 0$. But decreasing $R$, that is increasing $S$, the value of $\delta$ varies as a function of $S$. Thus in the plane $(S,\delta)$ we have a random walk which,
 according to \eqref{a0}, has  $<\delta(S)>=0, <\delta^2(S)>=S$ for any $S$. A walk which has been constructed by  simply choosing $\delta(S)$ from a distribution given by  \eqref{a0} has  uncorrelated steps. But this is not the case  of interest. We are interested in  imposing the correlation between steps as this is given by \eqref{cor}.\\
 A general way to do this is to use Cholesky decomposition  \cite{recipes}.\\ An one column vector $\boldsymbol{\delta'}$ with components  $\delta'_0,\delta'_1...\delta'_n$ which correspond to $S_0,S_1...S_n$ with $S_0<S_1<...<S_n$  are chosen from a a central normal distribution of unit variance, $N(0,1)$. The correlation matrix for the values of $\boldsymbol{\delta}'$ is $E[\boldsymbol{\delta}'\boldsymbol{\delta'}^T]=\boldsymbol{I}_{n+1}$, where $E$ denotes the expected value. The desirable correlation matrix \textbf{C} is given by
 \begin{equation}
 \textbf{C}=[<\delta_i\delta_j>], i,j=0,n
 \end{equation}
 which is a symmetric and positive definite with its elements  given by \eqref{cor}, is decomposed and  written as
 \begin{equation}
 \textbf{C}=\textbf{L}\textbf{L}^T
 \end{equation}
 Then the vector $\boldsymbol{\delta}=\boldsymbol{L}^T\boldsymbol{\delta'}$ has the desirable correlation between its components. The proof is simple. The correlation matrix for the values of $\boldsymbol{\delta}$ is $E[\boldsymbol{\delta}\boldsymbol{\delta}^T]$ which can be written as
 \begin{gather}
 E[\boldsymbol{\delta}\boldsymbol{\delta}^T]=E[\boldsymbol{L\delta'}(\boldsymbol{L\delta}')^T]=
 E[\boldsymbol{L\delta'}\boldsymbol{{\delta'}^TL^T}]=\boldsymbol{L} E[\boldsymbol{\delta'}\boldsymbol{\delta'}^T]\boldsymbol{L}^T=\nonumber\\
 \boldsymbol{LI_{n+1}L^T}=\boldsymbol{LL^T}=\boldsymbol{C}
 \end{gather}
 For the case, in this paper \cite{Hiotelis2017} proposed we study the stochastic process of the form
 \begin{equation}
 \label{b1}
 \delta(S)=\int_{0}^S K(S,u)\mathrm{d}W(u)
 \end{equation}
 where $K$ is a kernel and $\mathrm{d}W$ is the usual Wiener process with the kernel $K$  given by
 \begin{equation}
 \label{b2}
 K(S,u)=c\left[1-a\frac{u}{S}\right]
 \end{equation}
 for $u\leq S$ and zero otherwise. This is a non-Markovian process. Under the conditions $<\delta^2(S)>=S$ and $c^2\left[\frac{a^2}{3}-a+1\right]=1,$ the resulting two point correlation is given by \eqref{cor} with
 \begin{equation}
 \lambda=\frac{a(3-2a)}{2(a^2-3a+3)}
 \end{equation}
 In what follows, we used both methods. We will compare the results from these two different methods later.\\
 Going bach to the point $A$,  of interest we assume that at the first value of $R$ or equivalently $S$ the mean overdensity exceeds a threshold $\delta_c$, then a halo of mass $m$ with  $\sigma^2(m)=S$ is formed.
  Let now consider an elementary  element of mass $\Delta m$ and volume $\Delta V$ which is placed randomly. If $N(m)\mathrm{d}m$ is the number of haloes with masses in the range  $m, m+\mathrm{d}m$, then the typical volume of such a halo is $V=m/\rho_b$, where $\rho_b$ is the background density of the Universe. Thus, the total volume of these haloes is $N(m)V$ and the probability the elementary element is placed inside such a halo, $f(m)\mathrm{d}m$, is simply $N(m)V/V_T$, where $V_T$ is the total volume of the Universe.Thus
 \begin{equation}
 f(m)\mathrm{d}m=\frac{N(m)\mathrm{d}m}{V_T}V=n(m)\mathrm{d}mV=n(m)\frac{m}{\rho_b}\mathrm{d}m
 \end{equation}
 where $n(M)=N(m)/V_T$ is the number density of such haloes. Thus, the number density of haloes of mass in the range  $m, m+\mathrm{d}m$, is given by
 \begin{equation}
 n(m)\mathrm{d}m=\frac{\rho_b}{m}f(m)\mathrm{d}m
 \end{equation}
 The excursion set model gives the ability to calculate $f(m)$ in order to find the number density for the above equation. We consider a number $N_T$ of trajectories. For each one of them we calculate its positions at a set of values $S_0<S_1<...S_n$. This is done or using the Cholesky decomposition method or the stochastic process described above, \cite{Hiotelis2017}. Each trajectory represents an elementary mass element. Let denote by $N_c$ the number of trajectories which first cross the threshold $\delta_c$  between $S, S+\mathrm{d}S$. Then,  $N_c/N_T$ is the fraction of the total mass of the Universe which belongs to haloes of mass  between $m, m+\mathrm{d}m $ with $\sigma^2(m)=S$. Thus, $f(S)\mathrm{d}S=N_c/N$ and
\begin{equation}
\label{mf1}
n(m)\mathrm{d}m=-\frac{\rho_b}{m}f(S)\mathrm{d}S=\frac{\rho_b}{m}f(S)\frac{\mathrm{d}S}{\mathrm{d}m}\mathrm{d}m
\end{equation}
which can be written as
\begin{equation}
\label{mf2}
n(m)\mathrm{d}m=\frac{\rho_b}{m^2}[2Sf(S)]\frac{\mathrm{d}\ln S^{-\frac{1}{2}}}{\mathrm{d}\ln m}\mathrm{d}m
\end{equation}
The quantity $2Sf(S)$  is called the multiplicity function. The rest of the terms in the above equation depend on the cosmological model and of course on the power spectrum.

\section{Results}
\subsection{The Monte Carlo approach and tests}
We divide the interval $[M_{min},M_{max}]=[10^{-3},10^{5.5}]M_{unit}$ of masses into $N_I$ intervals of equal length in logarithm spacing.  $N_T$ trajectories are considered and $\delta_i, i=1,2..N_I $ values with respective variances $S(M_{i}), i=1,2...N_I$ and correlations given by \eqref{cor}, are calculated either using the Cholesky decomposition method or the above described stochastic process. These trajectories are used to produce our results below.\\
In order to check the efficiency of the Monte Carlo method used, we first calculated multiplicity function and we compared with  approximations existing in the literature. In \cite{Maggiore2010a}, the authors use a path integral approach to estimate the first crossing distributions and their results have the form of infinite series which converge slowly.The resulting multiplicity function  is approximated by the formula,
\begin{equation}
\label{c2}
2Sf(S)=(1-\lambda)\left(\frac{2}{\pi}\right)^{1/2}\nu e^{-\frac{1}{2}{\nu}^2}+\frac{\lambda} {\sqrt{2\pi}} \nu G \left (\frac{1}{2} {\nu}^2\right)
\end{equation}
where $G(x)=\int_{x}^{\infty}t^{-1}e^{-t}\mathrm{d}t$, (see Eq. 120 in \cite{Maggiore2010a}) and $\nu=\delta_c(z)/\sqrt(S)$.\\
 We recall that $\delta_c(z)$ is the linear extrapolation up to present of the overdensity of a spherical region which collapses at redshift $z$, \cite{Peebles1980}. The number $n_i$ of trajectories which have their first upcrossing of the barrier between $S_{i-1}-S{_i}$ are grouped and the first crossing  distribution is calculated by $f(S_i)=n_i/[N(S_{i-1}-S{_i})]$. Finally, the multiplicity function is calculated by $2S_if(S_i)$.In Fig.1, we show multiplicity functions derived by our Monte Carlo approach as well as predictions of \eqref{c2}. All predictions are for $\lambda=0$ and $z=0$. The thick solid line shows the predictions of \eqref{c2}, open squares show the predctions of the stochastic process described above and the thin solid line shows the results derived by using the Cholesky decomposition. We have used $N_I=400$ and $N_T=10^6$.The agreement is very satisfactory.
 \begin{figure}
 \hspace{-0.5cm}
\includegraphics[width=10cm ]{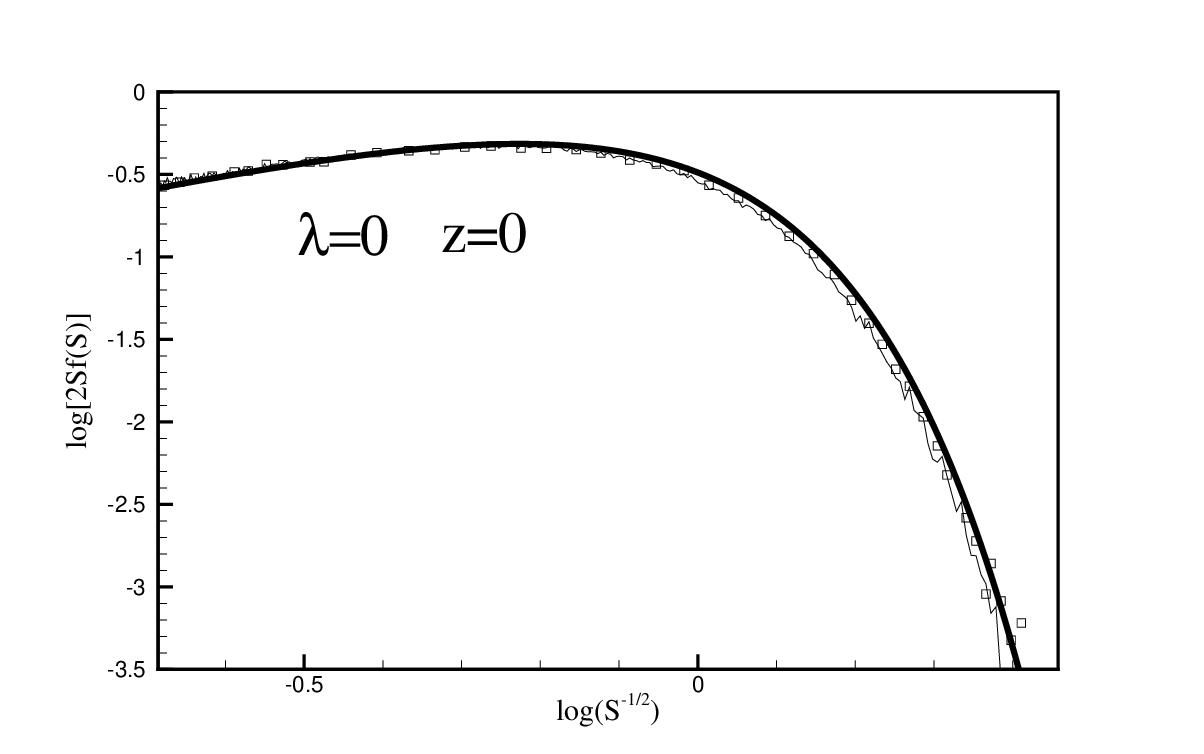}
\caption{ Multiplicity functions for z=0 and $\lambda=0$. The thick solid line is the predictions of the analytical approximation
of \cite{Maggiore2010a}. Empty squares are the predictions of the stochastic process and thin solid line the prediction of the Cholesky decomposition. Practically, the results are indistinguishable. }\label{fig1}
 \end{figure}
The resolution in mass, which depends on $N_I$, as well as the number of trajectories $N_T$, are crucial for the main purpose of this paper which, as it is mentioned in the introduction, is to study effects on the number density of haloes due to their environment. In Fig.2, we plot multiplicity functions from the two methods for $\lambda=0.45$ and $z=2$ but for a considerably coarser resolution that is for $N_I=30$. Thick solid line shows the results of \eqref{c2}, open squares the results of the stochastic process and the thin solid line the results of the Cholesky decomposition. The last method has problems to agree with the analytical formula due to the small value of $N_I$. Thus, in what follows, we used the method of the stochastic process. In any of the cases which will study  the agreement of the multiplicity functions from the Monte Carlo method and those from \eqref{c2} has been checked and found satisfactory.
In Fig.3, the smooth dotted line is the prediction of \eqref{c2} while the crooked solid line is the prediction of our Monte carlo approximation.\\
We note, at this point, that the multiplicity functions which result from N-body simulations do not fit the predictions of \eqref{c2}. It is noted in \cite{Grossi2009} and in \cite{Hiotelis2017}  that an excellent fit is achieved when the usual threshold of collapse $\delta_c(z)$ is replaced by a new lower threshold $\delta^{*}_c(z)\equiv p\delta_c(z)$ where $p=0.866$. Before comparing  our results with those of \eqref{c2} as well as with those  of N-body simulations, it is  worthmentioning  here that a large number of fitting formulae of multiplicity functions is present in the literature, among. which are the following

 \begin{figure}
 \hspace{-0.5cm}
 \includegraphics[width=10cm]{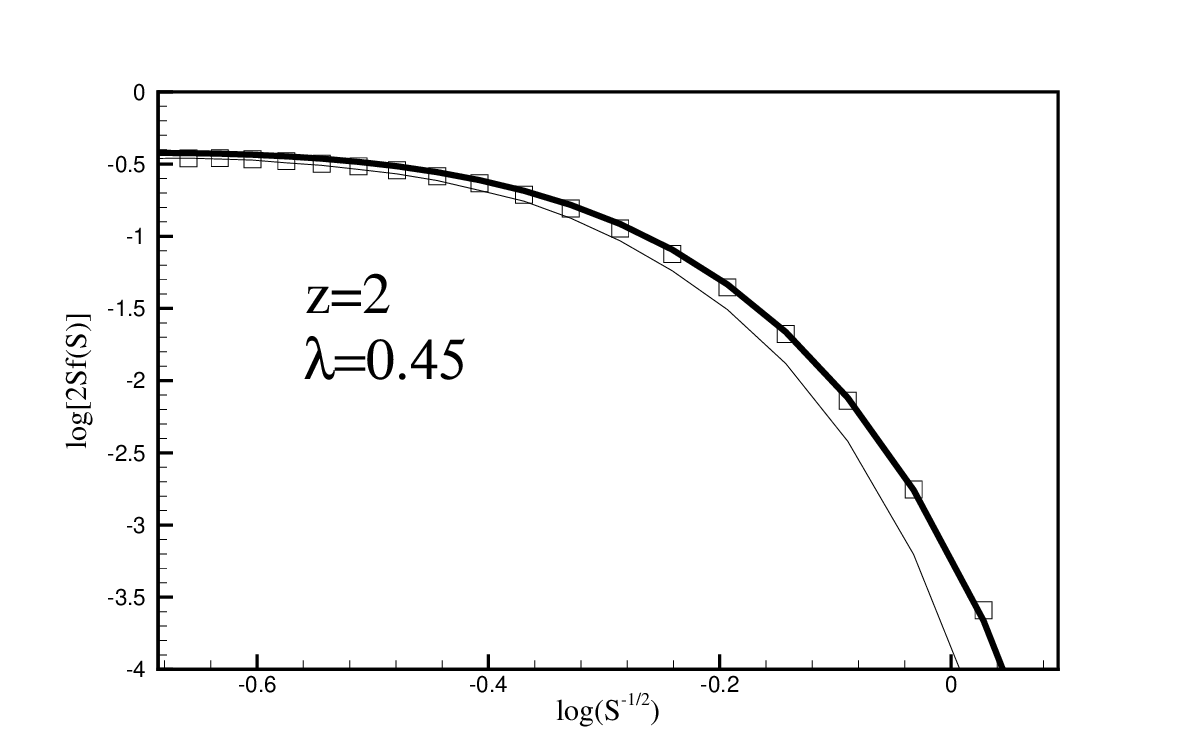}
\caption{ A comparison of the results of our Monte Carlo methods with the results of the formula \eqref{c2}. Open squares shows the prediction of the stochastic process described in \cite{Hiotelis2017} given by
\eqref{b2}. Thick solid line gives the results of formula \eqref{c2} and the thin solid line the predictions using the Cholesky decomposition. In all cases $z=2$, $\lambda=0.45$ and $N_T=30$. We see that the Cholesky decomposition is sensitive to the number of grid points $N_T$. Thus, in what follows we use the stochastic process \eqref{b2}.}\label{fig2}
 \end{figure}

 \begin{figure}
 \hspace{-0.5cm}
 \includegraphics[width=10cm]{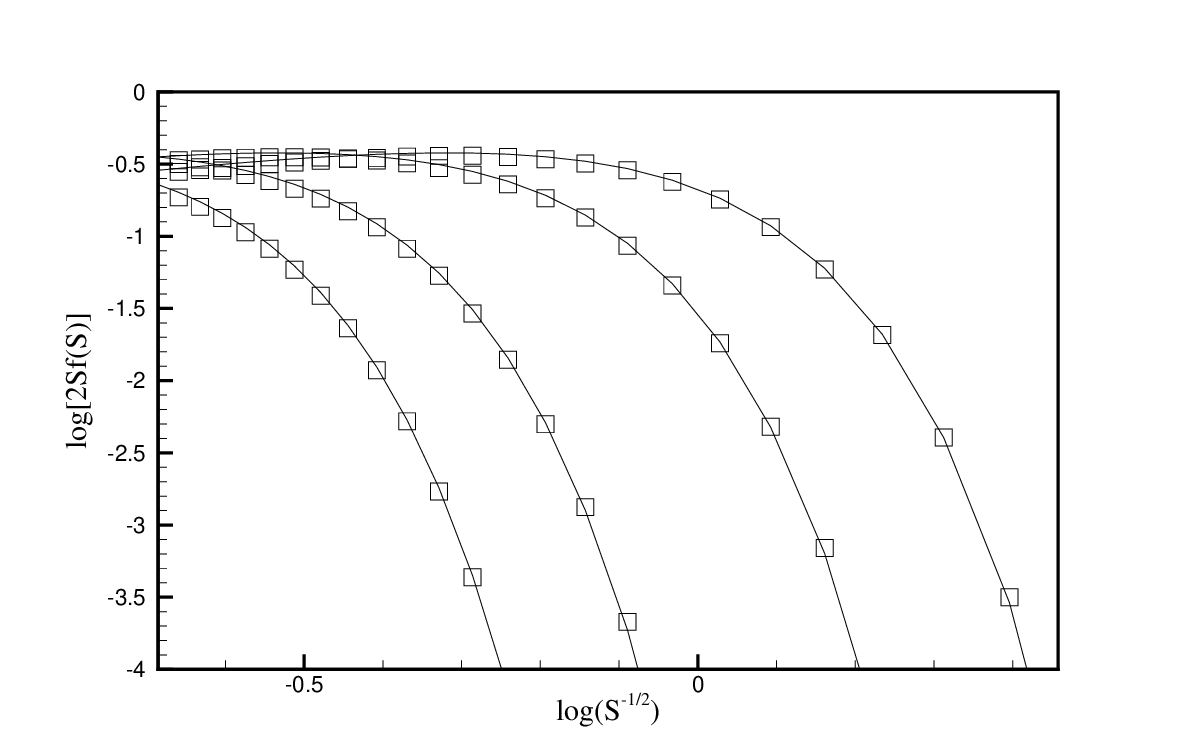}
\caption{ Multiplicity functions from the analytical formula of \eqref{c2} are represented by solid lines. Squares are the predictions of the stochastic processes. Curves from to the top to the bottom correspond to redshifts $z=0$,$z=1$, $z=3$ and $z=5$ respectively. The results of Monte Carlo are predicted for $N_I=30$ and $N_T=10^7$ and for $\lambda=0.45$. The agreement is very satisfactory. }\label{fig3}
 \end{figure}
 \begin{figure}
 \includegraphics[width=10cm]{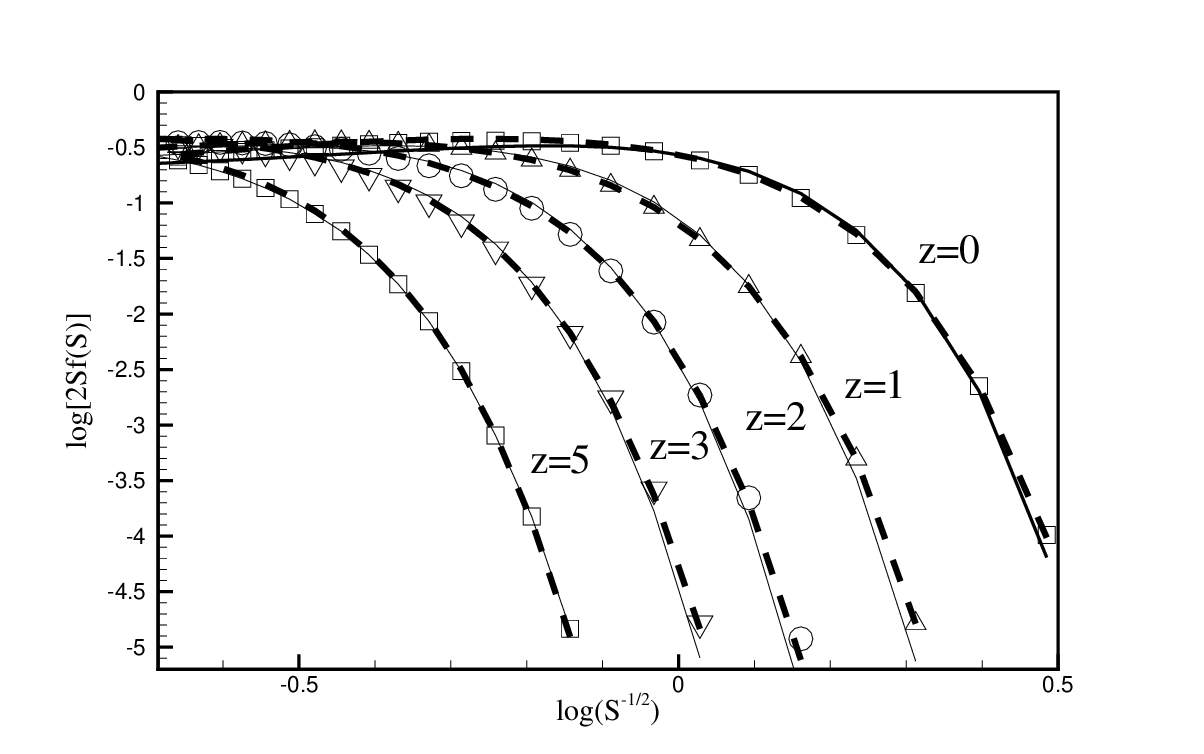}
 \hspace{-2.5cm}
\caption{ Empty symbols are the predictions of our model, thick dashes are the predictions of  the model of \cite{Maggiore2010a} and the solid line shows the results of the fitting formula of \cite{Warren2006} given by \eqref{war}. The results are predicted for $N_I=30$ and $N_T=10^7$.}\label{fig4}
 \end{figure}
 The fitting formula of \cite{Tinker2008} (given in their Eq.3)  which is

 \begin{equation}
 \label{tinkform}
 MF_T(\sigma,z)=A\left[\left(\frac{b}{\sigma}\right)^{a}+1\right]e^{-\frac{c}{\sigma^2}}
 \end{equation}
 where  $\sigma=\sqrt{S}$ and the z-dependence is given by
 \begin{gather}
 A=0.186(1+z)^{-0.14},~~
 a=1.47(1+z)^{-0.06},\nonumber\\
 b=2.57(1+z)^{-\alpha},~~
 c=1.19
 \end{gather}
 \begin{equation}
 \label{tinkforma}
  \log\alpha=-\left(\frac{0.75}{\log(\Delta_{vir}/75)}\right)^{1.2}
 \end{equation}
 ( See Eqs 5,6,7 and 8 in  \cite{Tinker2008}) where  $\Delta_{vir}\approx 200$. \\
  The formula of   \cite{Watson2013} which is valid for $0\leq z \leq 30$. This is given by
 \begin{equation}
 MF_{Wats}=MF_T
 \end{equation}
 where
 \begin{equation}
 A=0.282,~~ a=2.163,~~ b=1.406,~~ c=1.21
 \end{equation}
 The formula of \cite{Warren2006}, which is,
 \begin{equation}
 \label{war}
 MF_{War}=0.7234(\sigma^{-1.625}+0.2538)e^{-\frac{1.1982}{\sigma^2}}
 \end{equation}
 and the  formula of \cite{Sheth2001}
 \begin{equation}
 MF_{ST}=A\sqrt{\frac{2a_s}{\pi}}\left[1+\left(\frac{\sigma^2}{a_s\delta_c^2}\right)^{p_{s}}\right]
 \frac{\delta_c}{\sigma}e^{-\frac{a_s\delta_c^2}{2\sigma^2}}
 \end{equation}
 where $A=0.3222,~a_s=0.707$ and $p_s=0.3$.\\
 In Fig.4 we compare the predictions of \eqref{c2} with those of the stochastic process \eqref{b1} and with those of the analytical formula of
 \eqref{war} at various redshifts. Empty symbols depict the predictions of our model, thick dashes are the predictions of  the model of \cite{Maggiore2010a} and the solid line shows the results of the fitting formula of \cite{Warren2006} given by \eqref{war}. The agreement is very satisfactory so in what follows we use the threshold  $\delta^{*}_c$.
 \subsection{The environmental dependence}
 As in \cite{Zhang2014},we use indexes $p$, $d$ and $e$ to declare progenitors, descendants haloes and the environment respectively. The corresponding values of $S$ are $S_p>S_d>S_e$ and the masses are $m_p<m_d<m_e$. Additionally, we consider the redshifts $z_p>z_d$ and the corresponding values of the threshold $\delta_{cp}=\delta_c(z_p)$ and $\delta_{cd}=\delta_c(z_d)$. Since $z_p>z_d$ we have $\delta_{cp}> \delta_{cp}$. An illustration is shown in Fig.5.\\
 Let $N(S_e, \delta_e)$ is the number of trajectories which at $S_e$ lie in between $\delta_e$ and $\delta_e+\mathrm{d}\delta_e$. Additionally let $N(S_d,z_d;S_e,\delta_e)$ be the number of trajectories which have their first upcrossing of the barrier $\delta_{cd}=\delta(z_d)$  between $S_d, S_d+\mathrm{d}S_d$ and  at $S=S_e$ lie  between $\delta_e$ and $\delta_e+\mathrm{d}\delta_e$. Similarly, $N(S_p, z_p; S_d,z_d;S_e,\delta_e)$ is the number of trajectories which have their first upcrossing  of a barrier $\delta_{zp}$  between $S_p, S_p+\mathrm{d}S_p$, their first upcrossing of a barrier
  $\delta_{cd}=\delta(z_d)$  between $S_d, S_d+\mathrm{d}S_d$ and at $S=S_e$ lie  between $\delta_e$ and $\delta_e+\mathrm{d}\delta_e$. Then, from the two equations that follow
  \begin{equation}
  \label{e1}
  P(S_d,z_d/S_e,\delta_e)\mathrm{d}S_d=\frac{N(S_d,z_d;S_e,\delta_e)}{N(S_e,\delta_e)}
  \end{equation}
  and
  \begin{equation}
   \label{e2}
  P(S_p, z_p / S_d,z_d;S_e,\delta_e)\mathrm{d}S_p=\frac{N(S_d, z_p; S_d,z_d;S_e,\delta_e)}{N(S_d,z_d;S_e,\delta_e)}
  \end{equation}
  \begin{figure}
 \hspace{-0.5cm}
 \includegraphics[width=10cm]{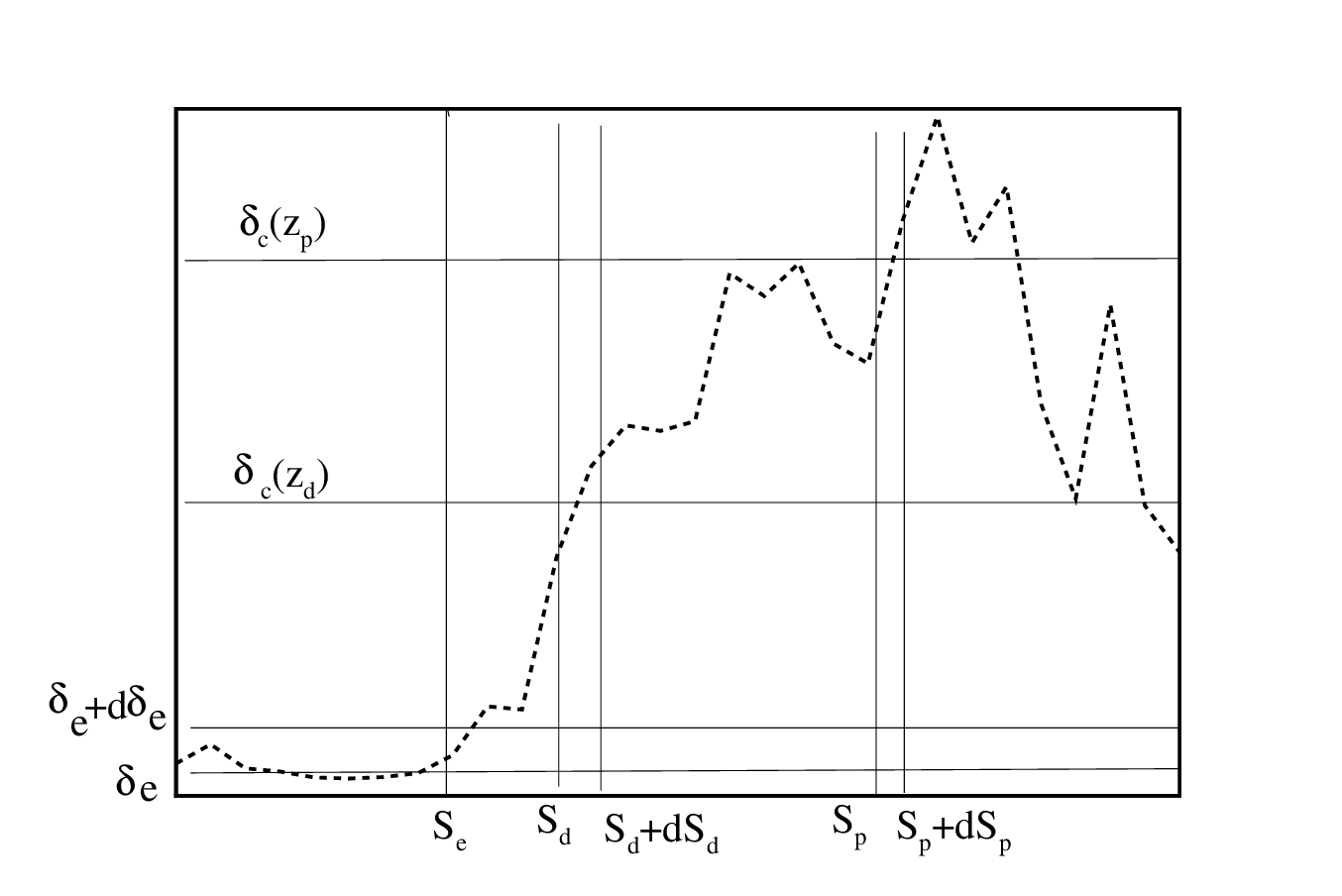}
\caption{ An illustration of the method used in order to calculate constrained probabilities used in our calculation. The figure shows a trajectory which at $S_e$ lies between $\delta_e$ and $\delta_e+\mathrm{d}\delta_e$ while it has its first upcrossing of the barrier $\delta_{cd}=\delta_c(z_d)$ between $S_d, S_d+\mathrm{d}S_d$  and its first upcrossing   of the higher barrier barrier $\delta_{cp}=\delta_c(z_p)$  between $S_p, S_p+\mathrm{d}S_p$.}\label{fig5}
 \end{figure}
  we find the probability of a trajectory having its first upcrossing of $\delta_{cp}$  between $S_p, S_p+\mathrm{d}S_p$  given that at $S_e$ is between $\delta_e$ and $\delta_e+\mathrm{d}\delta_e$ and the probability a trajectory have its first upcrossing of $\delta_{cp}$  between $S_p, S_p+\mathrm{d}S_p$  given that it has crossed the barrier $\delta_{cd}$  between $S_d, S_d+\mathrm{d}S_d$ and at $S_e$ is between $\delta_e$ and $\delta_e+\mathrm{d}\delta_e$.
  Then, the number of haloes is calculated by
  \begin{equation}
   \label{e3}
  N(m_d,z_d/m_e,\delta_e)=-\frac{m_e}{m_d}P(S_d,z_d/S_e,\delta_e)\frac{\mathrm{d}S_d}{\mathrm{d}m_d}
  \end{equation}
  and
  \begin{equation}
   \label{e4}
  N(m_p,z_p/ m_d,z_d;m_e,\delta_e)=-\frac{m_d}{m_p}P(S_p,z_p/ S_d,z_d;S_e,\delta_e)\frac{\mathrm{d}S_p}{\mathrm{d}m_p}
  \end{equation}
  \begin{figure}
 \hspace{-0.5cm}
 \includegraphics[width=10cm]{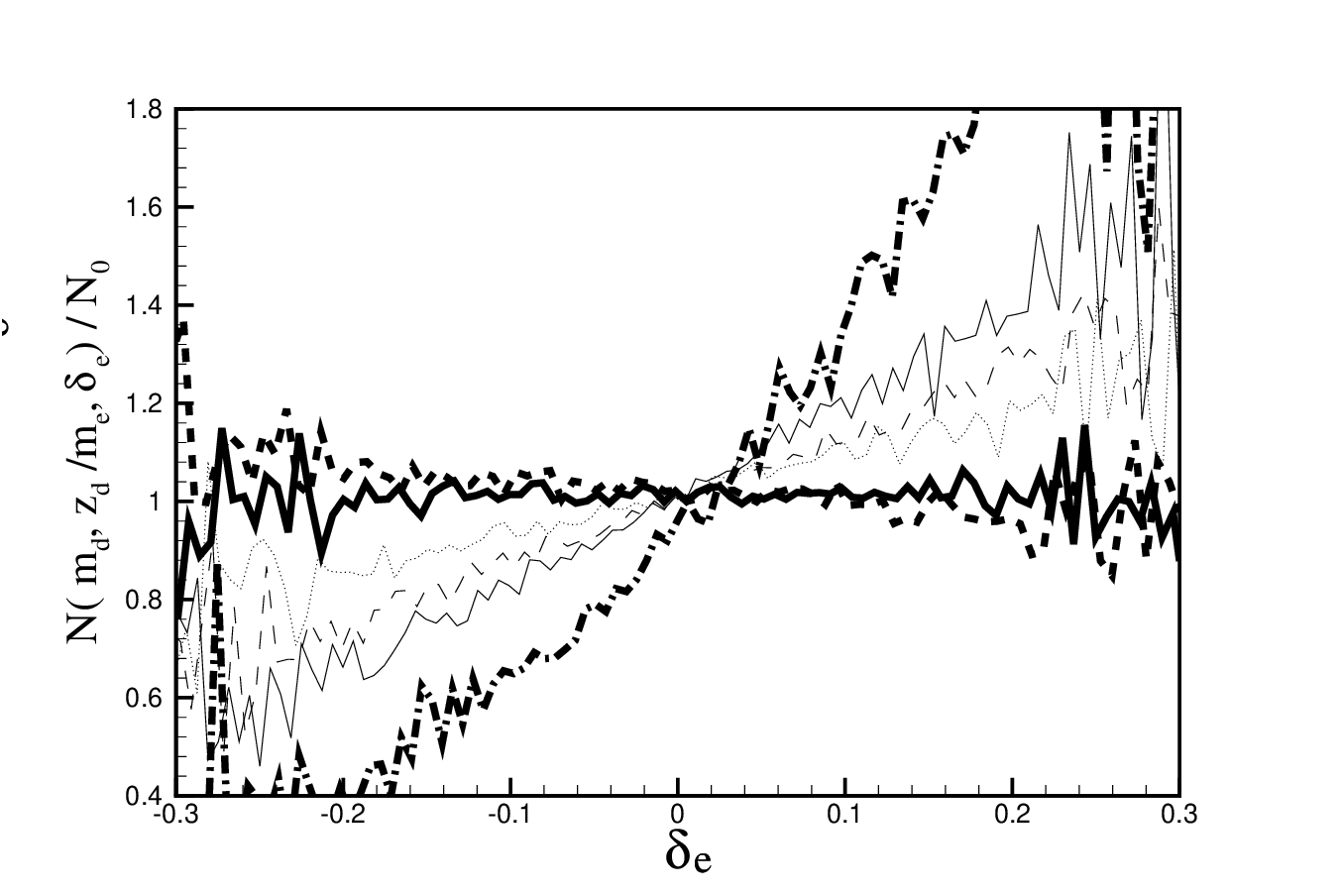}
 \caption{ Thick dashes  correspond to $m_d=8\times10^{9}M_{\odot}h^{-1}$, thick solid line depicts the results for $m_d=1.3\times10^{12}M_{\odot}h^{-1}$, dotted line shows the results for   $m_d=5\times10^{13}M_{\odot}h^{-1}$. Thin dashes are the results for  $m_d=\times10^{14}M_{\odot}h^{-1}$ and the thin solid line corresponds $m_d=2.14\times10^{14}M_{\odot}h^{-1}$. Finally, the line with dots and thick dashes corresponds to $m_d=9.26\times10^{14}M_{\odot}h^{-1}$.  It is clear that the number of small descendant haloes is a decreasing function of $\delta_e$ while the number of larger haloes increases with increasing $\delta_e$.The slope depends on the mass $m_d$. All curves have been normalized to the values $N(m_d,z_d/m_e, \delta_e=0)$}
\label{fig6}
 \end{figure}
 \begin{figure}
 \hspace{-0.5cm}
 \includegraphics[width=10cm]{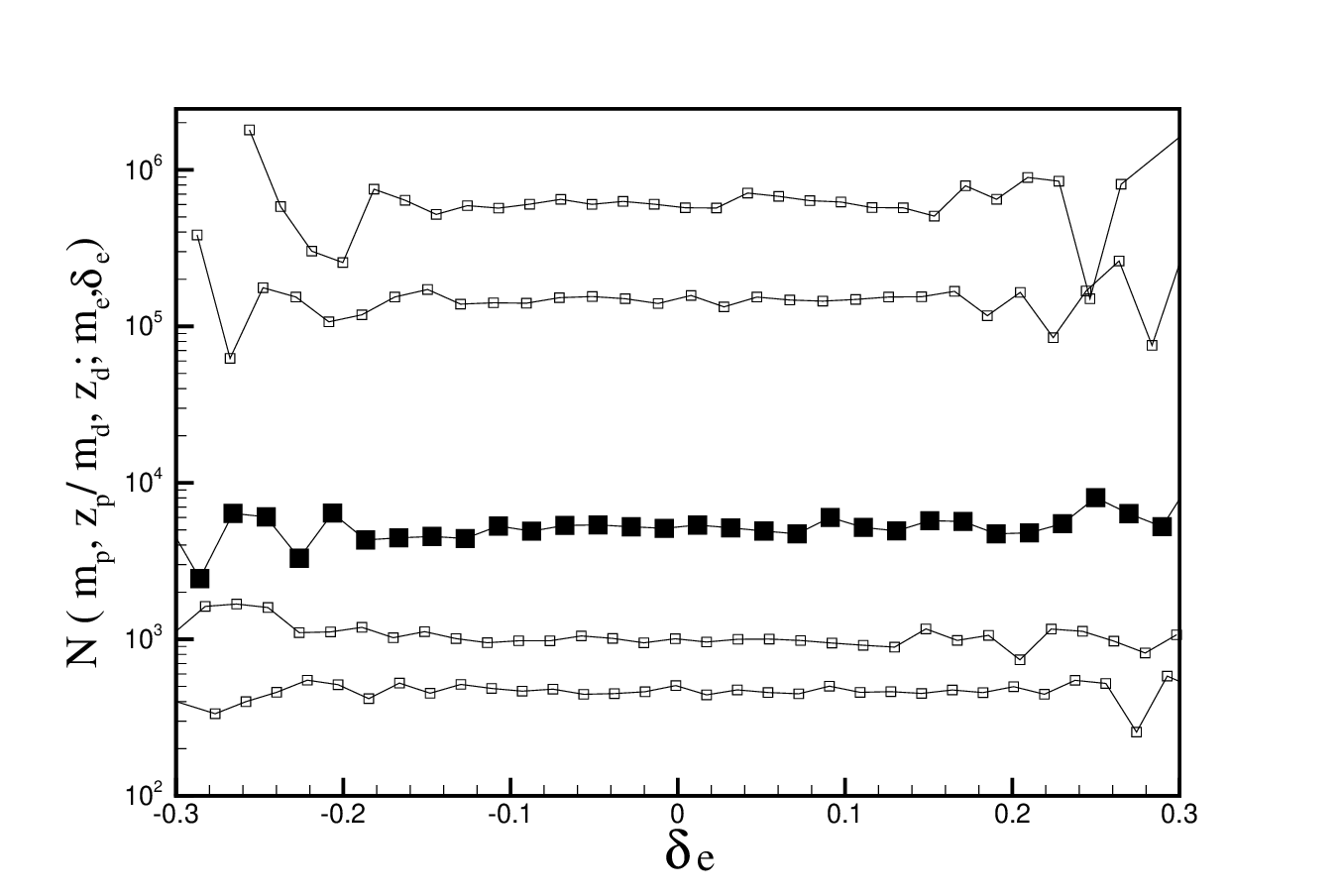}
 \caption{ In all cases  $m_p=1.85\times10^{9}M_{\odot}h^{-1}$. From  bottom to  top the curves correspond to $m_d=3.45\times10^{10}M_{\odot}h^{-1}$,
 $m_d=1.46\times10^{11}M_{\odot}h^{-1}$, $m_d=5.62\times10^{12}M_{\odot}h^{-1}$, $m_d=5.13\times10^{13}M_{\odot}h^{-1}$ and $m_d=2.15\times10^{14}M_{\odot}h^{-1}$. No dependence on $\delta_e$ is detected.}
\label{fig7}
 \end{figure}
  In Fig.5, an example of a trajectory which  has its  first upcrossing  of a barrier $\delta_{cd}=  \delta(z_d)$  between $S_d, S_d+\mathrm{d}S_d$, its  first upcrossing of a higher barrier  $\delta_{cp}=\delta(z_d)$ in between $S_p, S_p+\mathrm{d}S_p$ and at $S=S_e$ lies between $\delta_e$ and $\delta_e+\mathrm{d}\delta_e$ is shown.\\
  Our numerical scheme is as follows: First the positions of every of the $N_T$ trajectories at $S_e$ are stored. That is we store $(S_e, \delta_i) ,i=1,...N_T$. Then, we create two groups of trajectories. The first group $(G1)$ contains  those trajectories which satisfy the condition one $(C1)$, that is they have their first upcrossing of the lower barrier $\delta_{cd}$  between $S_d$ and $S_d+\mathrm{d}S_d$. The positions of these trajectories at $S_e$ are stored. The second group $(G2)$ contains those particles which satisfy $(C1)$  but, additionally,   have their first upcrossing of the higher  barrier $\delta_{cp}$  between $S_p$ and $S_p+\mathrm{d}S_p$, thus they satisfy $(C2)$.  Their positions at $S_e$ are also stored. The number of trajectories of every group is found. Then, for every group the minimum and the maximum value of $\delta$ at $S_e$ are found. The interval $[\delta_{min}, \delta_{max}]$ is divided into $n$ bins. We calculate the number of trajectories from the total number $N_T$ which belong to each bin. Additionally  the number of trajectories of each  group in every bin is  calculated and then the above equations \eqref{e1},\eqref{e2},\eqref{e3},\eqref{e4} are used.
  \begin{figure}
 \hspace{-0.5cm}
 \includegraphics[width=10cm]{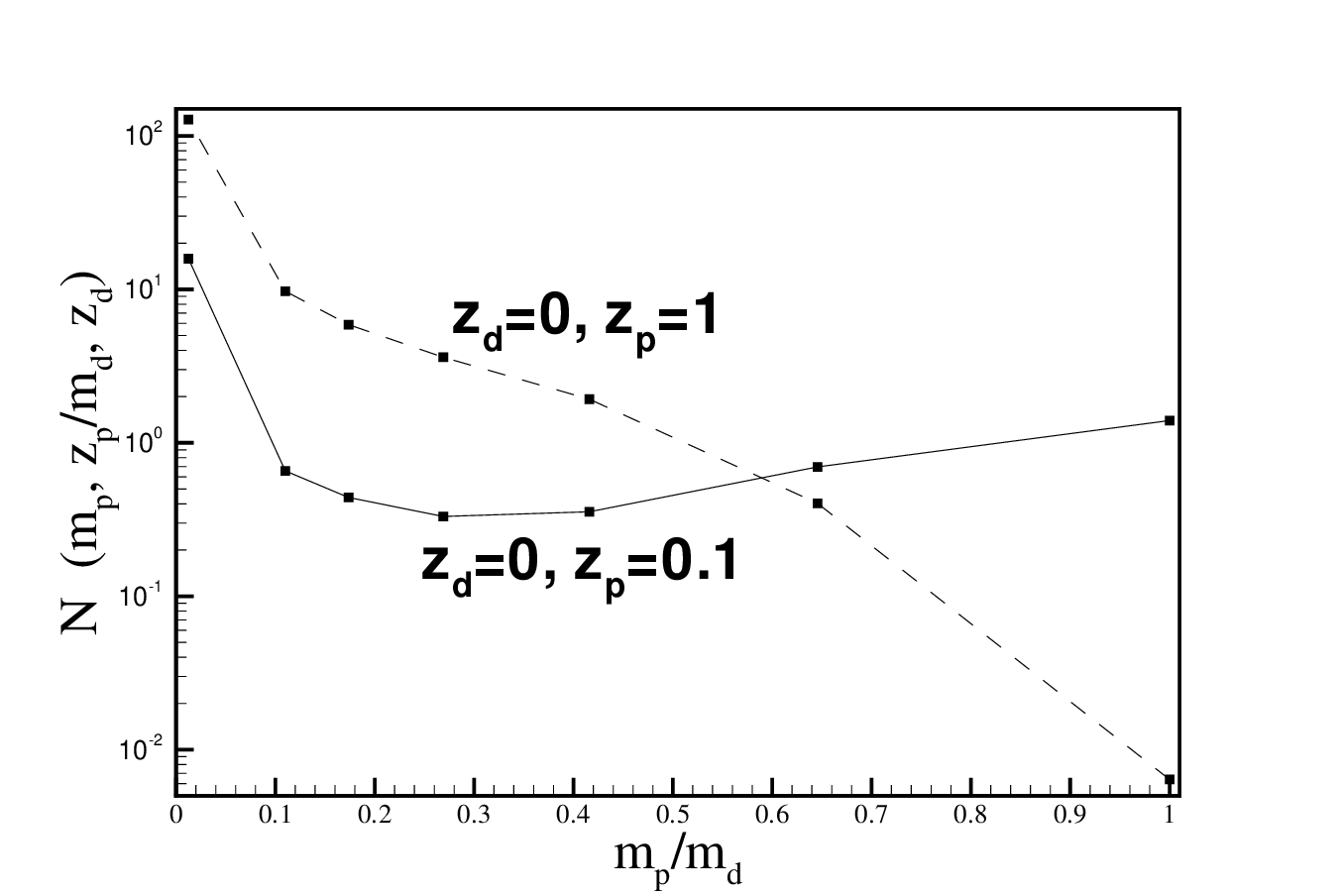}
 \caption{ The progenitors mass function of a present day halo, $z_d=0$, of mass   $m_d=1.3\times10^{12}M_{\odot}h^{-1}$.  Dotted line shows the distribution of the number of progenitors at $z_p=1$, while the solid line shows the  distribution at $z_p=0.1$}
 \label{fig8}
 \end{figure}

   We studied more than one hundred cases for different values of $N_I, N_T$,$z_p,z_d, m_p,m_p, m_e$. We present some of them which are characteristic and very helpful for understanding the whole mechanism of the assembly history of haloes. Thus, we can draw some conclusions about the environmental dependence of the formation history of haloes.\\
   First, we fixed $N_I=30$, $N_T=10^7$, $z_d=0$,  $z_p=1$, $m_e=1.5\times10^{17}M_{\odot}h^{-1}$ and we studied cases for the other parameters.
  In Fig.6 we show  $N(m_d,z_d/m_e,\delta_e)$ as a function of $\delta_e$. Thick dashes  correspond to $m_d=8\times10^{9}M_{\odot}h^{-1}$, dotted line depicts the results for $m_d=5\times10^{13}M_{\odot}h^{-1}$, thin dashes are the results for  $m_d=1\times10^{14}M_{\odot}h^{-1}$ and the solid line corresponds $m_d=2.14\times10^{14}M_{\odot}h^{-1}$. It is clear that the number of small descendant haloes is a decreasing function of $\delta_e$ while the number of larger haloes increases by increasing $\delta_e$. The slope depends on the mass $m_d$. All curves have been normalized to the value $N_0\equiv N(m_d,z_d/m_e, \delta_e=0)$.\\
  While the dependence of $N(m_d,z_d/m_e,\delta_e)$ on $\delta_e$ is clearly connected to the ratio $m_d/m_e$  no dependence of $N(m_p,z_p/ m_d,z_d;m_e,\delta_e)$ on  $\delta_e$ is observed. In Fig.7 we show  the quantity $N(m_p,z_p/ m_d,z_d;m_e,\delta_e)$ as a function of $\delta_e$ for various values of $m_d$. In all cases  $m_p=1.85\times10^{9}M_{\odot}h^{-1}$. From  bottom to  top the curves correspond to increasing mass of the progenitors $m_p$ with precise values which are written in figure caption.  No dependence of this number on $\delta_e$ is detected, since it is clear that is constant. We note that $N(m_p,z_p/ m_d,z_d;m_e,\delta_e)$ appears to be independent on $\delta_e$ in all cases that we have studied which are more than a hundred. Additionally we studied cases  using a larger number of trajectories and finer grid for mass range. The results are the same.\\
 Finally, in Fig.8, we give the total number of progenitors of a descendant halo of present day mass $m_d=1.3\times10^{12}M_{\odot}h^{-1}$ for two different redshifts of the progenitors, namely $z_p=1$ and $z_p=0.1$. As it is expected, the number of small progenitors decreases by decreasing redshift while the number of large progenitors increases with decreasing redshift. This is the expected behavior according to the picture of the hierarchical scenario of structure formation.\\
 \section{Discussion}
 Several ideas and conclusions of our study exist in the literature. A discussion about related interesting papers follows:\\
 I) In a pioneered paper \cite{Mo1996}  use an extension of the Press-Schechter formalism to study the statistical distribution of
 dark haloes in an initial Gaussian density field. They find that the bias function ($b(M,R,\delta)$) resulting from their model and defined by the relation $b(M,R,\delta)=b_h(M,R,\delta)/\delta$ where $b_h(M,R,\delta)$ is the mean overdensity of haloes of mass $M$ within spheres which have radiuw $R$ and mass overdensity $\delta$, describes accurately the results of N-body simulations. Thus   even uncorrelated walks model will predict a trend like this of Fig.6. This is a robust result and it can be verified by our for $\lambda =0$, but uncorrelated walks are not of interest here since they do not result to the correct multiplicity functions. It is very interesting  to note that the slopes in Fig.6 are close to those produced by a linear bias.\\
 II) Effects arising from the correlation between steps have also been studied in several important studies. In \cite{Musso2014} a model where correlations between steps arise because of nearest neighbor interactions. In their analysis of assembly bias ($4.3$, therein) the truncated kernels used are similar to our kernel. However, it is expected a similar assembly bias as it is clearly stated in our results.\\
 III) Finally, the height of the barrier has been studied in the literature  \cite{Sheth2001}, \cite{Ach2013}, \cite{Paranjape2013}, \cite{Maggiore2010b}. Since a main purpose of theoretical models is to produce observable characteristics that fit those of N-body simulations, the models studied above span from those which use a scale depended barrier (ellipsoidal collapse) for uncorrelated walks \cite{Sheth1999} to those using correlated walks and a barrier independent on scale \cite{Maggiore2010b}. The factor $p=0.866$ used in our calculations appears in the spherical part of the ellipsoidal collapse model of \cite{Sheth1999} and a justification, in terms of a diffusing barrier, is given in \cite{Maggiore2010b}.\\
 Our simple model is fully consistent with the idea of the excursion set approach, produces  a number of characteristics of dark matter haloes, in agreement with most of the results of other authors, and sheds more lights to underlying physical procedures during the formation of structure in the Universe.

\section{Acknowledgements}
\ We acknowledge  Dr. Chara Ganetsou for her kind help.


\begin{thebibliography}{}



\bibitem[Achitouv et al.(2013)]{Ach2013} Achitouv I., Rasera Y., Sheth R.~K., Corasantini P.~S., 2013,Physical Review Letters,111,231303

\bibitem[Bond et al. (1991)]{Bond1991} Bond J.R., Cole S., Efstathiou G., Kaiser
N., 1991, ApJ, 379, 440



\bibitem[Bower (1991)]{Bower1991} Bower, R.~G. 1991, \mnras, 248, 332











\bibitem[Gao \& White (2007)]{Gao2007}
Gao, L. \& {White}, S.~D.~M. 2007, \mnras,377, L5

\bibitem[Gardner(2001)]{Gardner2001}
Gardner, J.~P. 2001, \apj, 557, 616

\bibitem[{{Grossi} {et~al.}(2006){Grossi}, {Verde}, \&
  {Carbone}}]{Grossi2009}
{Grossi }, M., {Verde}, L., \& {Carbone}, K. 2009, \mnras, 398, 321









\bibitem[Hiotelis \& Del Popolo (2017)]{Hiotelis2017}
Hiotelis, N. \& Del Popolo, A. 2017, \aap, 607, A47

\bibitem[{{Jenkins} {et~al.}(2001){Jenkins}, {Frenk}, {White}, {Colberg},
  {Cole}, {Evrard}, {Couchman}, \& {Yoshida}}]{Jenkins2001}
{Jenkins}, A., {Frenk}, C.~S., {White}, S.~D.~M., {et~al.} 2001, \mnras, 321,
  372


\bibitem[{{Lacey} \& {Cole}(1993)}]{Lacey1993}
{Lacey}, C. \& {Cole}, S. 1993, \mnras, 262, 627


\bibitem[Maggiore \& Riotto(2010)]{Maggiore2010a}
{Maggiore}, M. \& {Riotto}, A. 2010, \apj, 711, 907

\bibitem[Maggiore \& Riotto(2010b)]{Maggiore2010b}
{Maggiore}, M. \& {Riotto}, A. 2010, \apj, 717, 515


\bibitem[Mo \$ White(1996)]{Mo1996}
Mo, H.~J., White, S.~D.~M., 1996, \mnras, 282,347



\bibitem[Musso \& Sheth (2014)] {Musso2014}
Musso, M., Sheth, R.~K., 2014,\mnras 443, 1601

\bibitem[Paranjape et. al. (2013)]{Paranjape2013}
Paranjape, A., Sheth R.K., Desjacques, V., 2013, \mnras, 431 1503

\bibitem[Peacock \& Heavens (1974)]{Peacock1990}
Peacock, J.~A. \& Heavens, A.~F., 1990, \mnras, 243,133

\bibitem[Peebles(1980)]{Peebles1980}
Peebles, P.~J.~E. 1980, {The large-scale structure of the universe}
  (Princeton University Press, Princeton)
\bibitem[Percival et~al.(2003)] {Percival2003}  Percival W.,J., Scott D., Peacock J.,A.,  \&
  Dunlop J.,S., 2003, \mnras, 338, L21


\bibitem[Press \& Schechter(1974)]{Press1974}
Press, W.~H. \& Schechter, P. 1974, \apj, 187, 425
\bibitem[Press et~al.(1990)]{recipes}
Press, W.~H.,Flannery, S.~A., Teukolsky, A., Vetterling, W.~T., {Numerical Recipes; The Art of Scientific Computing}
(Cambridge University Press, 1990)


\bibitem[Sheth et~al.(2001)]{Sheth2001}
Sheth, R.~K., Mo, H.~J., \& Tormen, G. 2001, \mnras, 323, 1

\bibitem[{{Sheth} \& {Tormen}(1999)}]{Sheth1999}
{Sheth}, R.~K. \& {Tormen}, G. 1999, \mnras, 308, 119

\bibitem[Sheth \& Tormen (2002)]{Sheth2004}
Sheth, R.~K. \& Tormen, G. 2004, \mnras, 350, 1383

\bibitem[Smith et~al.(1998)]{Smith1998}
{Smith}, C.~C., {Klypin}, A., {Gross}, M.~A.~K., {Primack}, J.~R., \&
 {Holtzman}, J. 1998, \mnras, 297, 910

\bibitem[Tinker et~al.(2008)]{Tinker2008}
{Tinker}, J., {Kravtsov}, A.~V., {Klypin}, A., {et~al.} 2008, \apj, 688, 709

\bibitem[Warren et~al.(2006)]{Warren2006}
{Warren}, M.~S., {Abazajian}, K., {Holz}, D.~E., \& {Teodoro}, L. 2006, \apj,
  646, 881

\bibitem[Watson et~al.(2013)]{Watson2013}
{Watson}, W.~A., {Iliev}, I.~T., {D'Aloisio}, A., {et~al.} 2013, \mnras, 433,
  1230


\bibitem[White(2002)]{White2002}
{White}, M. 2002, \apjs, 143, 241
\bibitem[Zhang et~al.(2014)]{Zhang2014}
Zhang, J., Ma, C-P., \& Riotto, A. 2014, \apj,
  782, 44
\bibitem[Zentner et~al.(2005)]{Zentner2005}
Zentner, A.~R., {Berlind}, A.~A., Bullock, J.~S., Kratsov, A.~V.,Weschsler, R.~H.,  2005, \mnras, \apj,
  624,505
\bibitem[Zentner {et~al.}(2014)]{Zentner2014}
Zentner, A.~R., Hearin, A.~P., \& van de Bosch, G.~C., 2001, \mnras, 323, 1
\end{thebibliography}
\end{document}